\makeatletter

\def\vereq#1#2{\lower3pt\vbox{\baselineskip1.5pt \lineskip1.5pt
\ialign{$\m@th#1\hfill##\hfil$\crcr#2\crcr\sim\crcr}}}

\def\gtrsim{\mathrel{\mathpalette\vereq>}}

\makeatother
\newcommand{\bm}[1]{\mbox{\boldmath$#1$}}
\documentclass[12pt]{iopart}
\begin{document}

\title[Three Dimensional Heisenberg Spin Glass Model]
{Three Dimensional Heisenberg Spin Glass Models 
with and without Random Anisotropy}

\author{F Matsubara\dag, T Shirakura\ddag, 
S Endoh\dag and S Takahashi\dag}

\address{\dag\
Department of Applied Physics, Tohoku University, Sendai 980-8579,
Japan}
\address{\ddag\
Faculty of Humanities and Social Sciences, Iwate University, 
Morioka 020-8550, Japan}

\ead{fumi@camp.apph.tohoku.ac.jp}

\begin{abstract}

We reexamine the spin glass (SG) phase transition of the $\pm J$ Heisenberg 
models with and without the random anisotropy $D$ in three dimensions 
($d = 3$) using complementary two methods, i.e., (i) the defect energy 
method and (ii) the Monte Carlo method. We reveal that the conventional 
defect energy method is not convincing and propose a new method 
which considers the stiffness of the lattice itself. 
Using the method, we show that the stiffness exponent $\theta$ has a 
positive value ($\theta > 0$) even when $D = 0$. 
Considering the stiffness at finite temperatures, we obtain the 
SG phase transition temperature of $T_{\rm SG} \sim 0.19J$ for $D = 0$. 
On the other hand, a large scale MC simulation shows that, in contrary to 
the previous results, a scaling plot of the SG susceptibility 
$\chi_{\rm SG}$ for $D = 0$ is obtained using almost the same transiton 
temperature of $T_{\rm SG} \sim 0.18J$. Hence we believe that 
the SG phase transition occurs in the Heisenberg SG model in $d = 3$. 

\end{abstract}

\input epsf


\maketitle

%%%%%%%%%%%%%%%%%%%%%%%%%%%%%%%%%%%%%%%%%%%%%%%%%%%%%%%%%%%%%%%%%%%%%%%%%%%%%

\section{Introduction}
For a long time, it has been believed that the spin glass (SG) phase is 
realized in three dimensions ($d =3$) for the 
Ising model\cite{Bhatt,Ogielski} but not for the XY and Heisenberg 
models.\cite{Banavar,McMillan1,Olive,Iyota1,Coluzzi}
Thus the SG phases observed in experiments were suggested to be realized 
due to anisotropy.\cite{BrayMY,Iyota2} 
However, numerical studies in the last decade have revealed that 
the SG phase might be more stable than what has been believed so far. 
In a long-range Ruderman-Kittel-Kasuya-Yoshida (RKKY) model, it was 
shown that the SG susceptibility exhibits a divergent singularity at a 
finite temperature, even when the anisotropy is absent.\cite{Iguchi,Iguchi2} 
This behavior has been attributed to the randomness of the spin position 
(site random model) rather than the long-range nature of the RKKY interaction. 
In fact, a short-range site random model composed of ferromagnetic spins 
and antiferromagnetic spins was shown to exhibit a long-range order 
phase characterized by the co-existence of a ferromagnetic 
and an antiferromagnetic orders.\cite{Tamiya} 
On the other hand, for the XY and Heisenberg bond SG models, Kawamura 
and his coworkers took notice of chiralities of the spins and showed that 
a chiral glass (CG) phase transition occurs at a finite temperature 
$T_{\rm CG} \neq 0$, but the spin glass phase is still 
absent.\cite{Kawamura2,Kawamura3,Kawamura4} They insisted that 
an anisotropy mixes the chiral freedom and the spin freedom 
and the SG phase transition occurs at $T_{\rm SG} (= T_{\rm CG})$. 
This view of the SG phase transition is quite attractive, 
because it gives a novel picture of the SG phase. 
That is, in the picture, the CG phase realizes in the real world, 
not the SG phase. 
However, their bases of the absence of the SG phase are obscure. 
Moreover, since the chirality is described by the spin variables, then the 
origin of the CG phase transition might be the usual SG phase transition. 
In fact, recent studies of the aging effects of the spin and the chirality 
autocorrelation functions\cite{Rot} and the developments of the SG and 
the CG susceptibilities\cite{Nakamura} by means of a nonequilibrium 
relaxation method suggested that, if the CG phase transition occurs,  
the SG phase transition occurs at the same transition temperature 
$T_{\rm SG} = T_{\rm CG}$. 
Quite recently, Lee and Young presented the same conclusion using 
a finite size analysis of the correlation length of the spins 
and chiralities.\cite{Lee} 

During the last decade, new algorithms for simulating the complex systems 
have developed and available computer power has increased enormously. 
It is therefore possible to reexamine in detail the SG phase transition 
of the Heisenberg model on the base of usual analyses.   
Here we consider Heisenberg models with and without random anisotropy on 
a simple cubic lattice described by 
\begin{equation} 
     H = - \sum_{\langle ij \rangle}[J_{ij}\bm{S}_{i}\bm{S}_{j}
    + \sum_{\alpha\neq\beta}D_{ij}^{\alpha\beta}S_i^\alpha S_j^\beta], 
\end{equation} 
where $\bm{S}_{i}$ is the Heisenberg spin of $|\bm{S}_i| = 1$ 
and $S_i^{\alpha}$ is its $\alpha$-component $(\alpha = x, y, z)$, 
and $\langle ij \rangle$ runs over all nearest-neighbor pairs. 
The exchange interaction $J_{ij}$ takes on either $+J$ or $-J$ with 
the same probability of 1/2. We assume that the anisotropy comes from
pseudo-dipolar couplings and impose the restriction
$D_{ij}^{\alpha\beta} = D_{ji}^{\alpha\beta} = D_{ij}^{\beta\alpha}$.
We further assume that $D_{ij}^{\alpha\beta}$ are uniform random values 
between $-D$ and $D$.

Evidences of the absence of the SG phase in the Heisenberg SG model 
which have been believed so far are the following two points.
%%%%%%%%%%%%%%%%%%%%%%%%%%%%%%%%%%%%%%%%%%%%%%%%%%%%%%%%%%%%%%%%%%%%%
\begin{enumerate}
\item Negative stiffness exponent $\theta$ at $T = 0$. 
       \cite{Banavar,McMillan1,Kawamura2}
\item Scaling plots of the SG susceptibility $\chi_{SG}$ and absence of 
      the crossing of the Binder ratio $g_L$.\cite{Olive,Kawamura5}
\end{enumerate}
%%%%%%%%%%%%%%%%%%%%%%%%%%%%%%%%%%%%%%%%%%%%%%%%%%%%%%%%%%%%%%%%%%%%%
Then we reexamine these two points to consider the possibility 
of the SG phase transition of the Heisenberg SG model. 
We will consider stiffness exponent $\theta$ at $T = 0$ and $T \neq 0$ 
in section 2, and properties of $\chi_{SG}$ and $g_L$ in section 3. 
We will give a special attention on the effect of the anisotropy, because 
it has been believed that the anisotropy brings the SG phase 
transition. So if it is true, we will find different properties between 
the models with and without the anisotropy.

%%%%%%%%%%%%%%%%%%%%%%%%%%%   Section 2   %%%%%%%%%%%%%%%%%%%%%%%%%%
\section{Stiffness exponent $\theta$}
The most accepted evidence of the absence of the SG phase is results of the 
defect energy method. So we first consider the defect energy method. 

\subsection{Conventional defect energy method} 
The defect energy method comes from an application of a 
renormalization-group idea.\cite{Banavar,McMillan1,BrayM} 
That is, one evaluates an effective coupling $\tilde{J}_L$ between 
block spins of the linear dimension $L$ generated by the renormalization. 
The effective coupling $\tilde{J}_L$ would depend on $L$ as 
$\tilde{J}_L \sim JL^{\theta}$ with $\theta$ being called as 
the stiffness exponent. When $\theta > 0$, the SG phase transition occurs 
at a finite temperature, while no phase transition occurs at any finite 
temperature when $\theta < 0$. 
To estimate $\tilde{J}_L$, one considers the domain wall energy 
$\Delta E_L$ which is defined as the difference in the ground state 
energy of two lattices A and B of size $L \times L \times L$ with the 
same bond distribution but with different boundary conditions. 
That is, for the lattice A, a periodic boundary condition is applied for every 
direction, and, for the lattice B, an antiperiodic boundary condition is 
applied for one direction and the periodic boundary condition 
for the other directions. By using the method, Banavar and Cieplak firstly 
estimated the value of $\theta \sim -1$ and predicted that the SG phase 
transition occurs at $T = 0$.\cite{Banavar} 
Successive studies also predicted negative values for $\theta$, i.e., 
$\theta \sim -0.65$\cite{McMillan1} and $\theta \sim -0.49$.\cite{Kawamura2} 

Recently, however, a doubt was thrown to the estimation of 
$\tilde{J}_L$.\cite{MSS1,KA} 
That is, in the calculation of $\Delta E_L$, one expects that 
no domain wall exists in the lattice A(or B) and hence one domain wall 
arises in the lattice B(or A). This expectation might be true, but 
another possibility would be equally true. That is, some domain wall will 
occur in the lattice A and some different domain wall in the lattice B. 
Then one might examine merely difference in the energy between those two 
domain walls. Does this defect energy $\Delta E_L$ really gives the effective 
coupling $\tilde{J}_L$ between the block spins? 
So we first examine $\Delta E_L$ of the model with and without 
random anisotropy. 
We calculate $\Delta E_L$ for lattices of $L \leq 9$, and for each $L$, 
the sample averages are taken over about 4000 
independent bond realizations. Results of $[|\Delta E_L|]$ 
are presented in Fig.1 in a log-log form, where $[ \cdots ]$ means 
the sample average. 
Data for $D = 0$ are curved. 
The most surprising thing is that this $L$ dependence of $[|\Delta E_L|]$ 
is similar to those in the case of $D \neq 0$. 
These results suggest two possibilities. 
One is that, a finite size effect is so large that the asymptotic region 
has not yet been reached. 
Since the curvature is upwards, it is possible that $\theta \sim 0$ or 
even more $\theta > 0$ in the limit of $L \rightarrow \infty$. 
The other is the inadequateness for estimating the defect energy 
as pointed out above. 
In order to examine the latter possibility, we study this problem 
in a different method.

%%%%%%%%%%%%%%%%%%%%%%%%%%%%%%%%%%%%%%%%%%%%%%%%%%% 
\begin{figure}
\epsfxsize=0.7\linewidth
\begin{center}
  \epsfbox{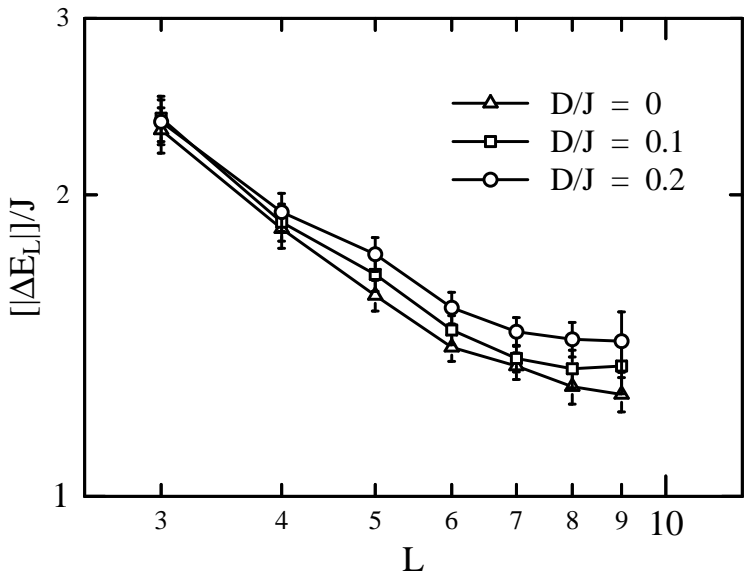}
\end{center}
  \caption{The lattice size dependence of the naive defect energy 
           $[|\Delta{E}_L|]$ of the Heisenberg SG model in $d = 3$.}
  \label{fig:1}
%\end{center}
\end{figure}
%%%%%%%%%%%%%%%%%%%%%%%%%%%%%%%%%%%%%%%%%%%%%%%%%%%%

\subsection{Stiffness of the system } 

Apart from the renormalization-group concept, we consider the stability 
of the spin configuration of the system itself.\cite{MSS1,MHZ} The strategy 
of our examination is as follows.\cite{MSS1,Endoh1,Shiomi,Endoh2} 
We prepare a cubic lattice of $L \times L \times (L+1)$ with an open 
boundary condition in one direction of $(L+1)$ lattice sites (z-direction) 
and the periodic boundary conditions in the other directions. 
That is, the lattice has two surfaces ${\Omega}_1$ and ${\Omega}_{L+1}$. 
We call this system the reference system. 
First we determine the ground state spin configuration of 
the reference system. 
Hereafter, the ground state spin configurations on ${\Omega}_1$ and 
${\Omega}_{L+1}$ are denoted as $\{{{\bm{S}}_i}^{(1)}\}$ and 
$\{{{\bm{S}}_i}^{(L+1)}\}$, respectively. 
In this spin configuration, any distortion (domain-wall) in the 
z-direction will be removed, because the lattice has free surfaces 
${\Omega}_1$ and ${\Omega}_{L+1}$. 
Then we add a distortion inside the system in the manner that 
$\{{{\bm{S}}_i}^{(1)}\}$ are fixed and $\{{{\bm{S}}_i}^{(L+1)}\}$ are changed 
under the condition that the relative angles between the spins are fixed. 
The ground state energy of this system is always higher than 
that of the reference system. 
When $D = 0$, this excess energy is the net one added inside the reference 
system, because the surface energy of ${\Omega}_{L+1}$, 
which is given as the sum of the exchange energies between 
the spins on ${\Omega}_{L+1}$, is conserved.
We consider the stability of the system on the basis of this excess energy. 
One might think that the fixing of the relative spin directions on 
${\Omega}_1$ and ${\Omega}_{L+1}$ overestimates the stability of the spin 
configuration. 
We think, however, that this restriction is not serious for discussing 
the stability, because the increase of the excess energy to infinity for 
$L \to \infty$ means nothing but the existence of a strong correlation 
between the spin configurations on ${\Omega}_1$ and ${\Omega}_{L+1}$. 
 In fact, the same method was successfully applied to the Ising SG model 
in $d = 2$.~\cite{MSS1,Shiomi}

We calculate two kinds of excess energies. One is the excess energy 
which is gained by rotating $\{{{\bm{S}}_i}^{(L+1)}\}$ by the same 
angle $\phi$ around some common axis (z-axis) and the other 
is the excess energy which is gained by reversing 
$\{{{\bm{S}}_i}^{(L+1)}\}$. Hereafter, we call the former 
system the rotated system and the latter system the reversed system.  
We think that it is sufficient to examine these two excess energies for 
considering the stiffness, because we can change $\{{{\bm{S}}_i}^{(L+1)}\}$ 
into any direction by combining the rotation and the reversal. 
The excess energy for the rotation $\Delta{E}_{\rm rot}(\phi)$ and that 
for the reversion $\Delta{E}_{\rm rev}$ are given as
\begin{eqnarray}
  \Delta{E}_{\rm rot}(\phi) &=&  E_{\rm rot}(\phi) - E_G ,\\
  \Delta{E}_{\rm rev} &=&  E_{\rm rev} - E_G ,
\end{eqnarray}
where $E_G$ is the ground state energy of the reference system, 
and $E_{\rm rot}(\phi)$ and $E_{\rm rev}$ are the ground state 
energies of the rotated system and reversed system, respectively. 
The lattice sizes studied here are $L =~3 - 8$ and, for each $L$, the 
sample averages are taken over about 1000 independent bond realizations.

In Fig.2, we present $L$-dependence of $[\Delta E_{\rm rot}(\pi/2)]$. 
Here we show data only for $D = 0$, because in the case of $D \neq 0$ 
we could hardly evaluate the net excess energy of 
$[\Delta E_{\rm rot}(\pi/2)]$.\cite{CommIsing} 
We clearly see that the data increase with $L$. 
From the slope of the asymptotic line shown in the figure, we tentatively 
determine the value of the stiffness exponent as $\theta_{\rm rot} \sim 0.62$.  That is, the SG phase would not be destroyed by a rotational perturbation.

In Fig.3, we present $L$-dependences of $[\Delta E_{\rm rev}]$ 
for both $D = 0$ and $D \neq 0$.  
Data depend little on the value $D$. 
For each of $D$'s, they seem to lie on a curve with a common positive 
slope of $\theta_{\rm rev} \sim 0.4$. 
Again, we get the view that the SG phase is stable at $T \ne 0$ for both 
$D \ne 0$ and $D = 0$. 

%%%%%%%%%%%%%%%%%%%%%%%%%%%%%%%%%%%%%%%%%%%%%%%%%
\begin{figure}
\epsfxsize=0.8\linewidth
\begin{center}
  \epsfbox{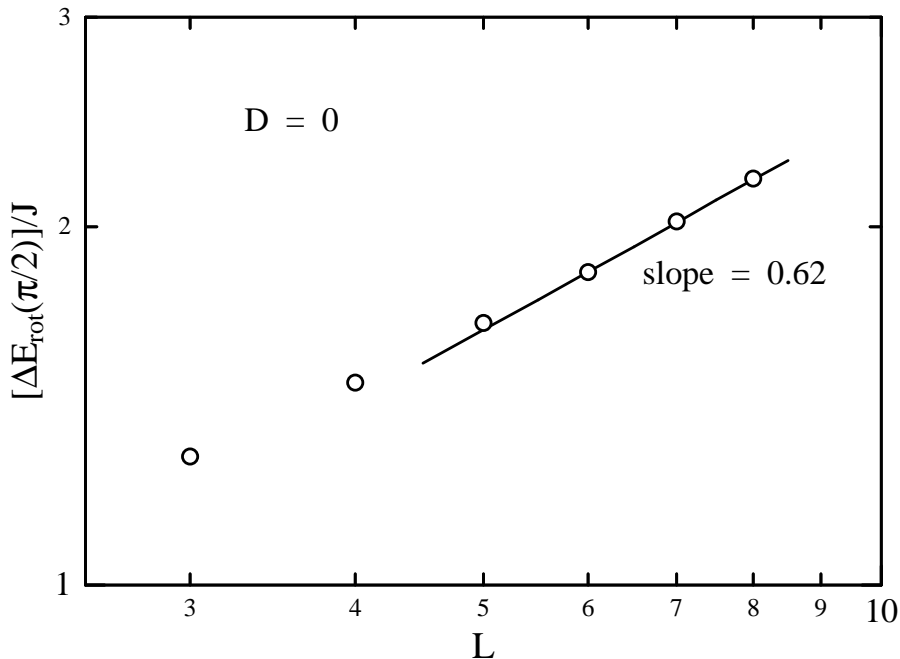}
\end{center}
  \caption{The lattice size dependence of the excess energy 
           $[\Delta{E}_{\rm rot}(\pi/2)]$ of the Heisenberg SG model in $d = 3$.}
  \label{fig:2}
\end{figure}
%%%%%%%%%%%%%%%%%%%%%%%%%%%%%%%%%%%%%%%%%%%%%%%%%
%%%%%%%%%%%%%%%%%%%%%%%%%%%%%%%%%%%%%%%%%%%%%%%%%
\begin{figure}
\epsfxsize=0.8\linewidth
\begin{center}
  \epsfbox{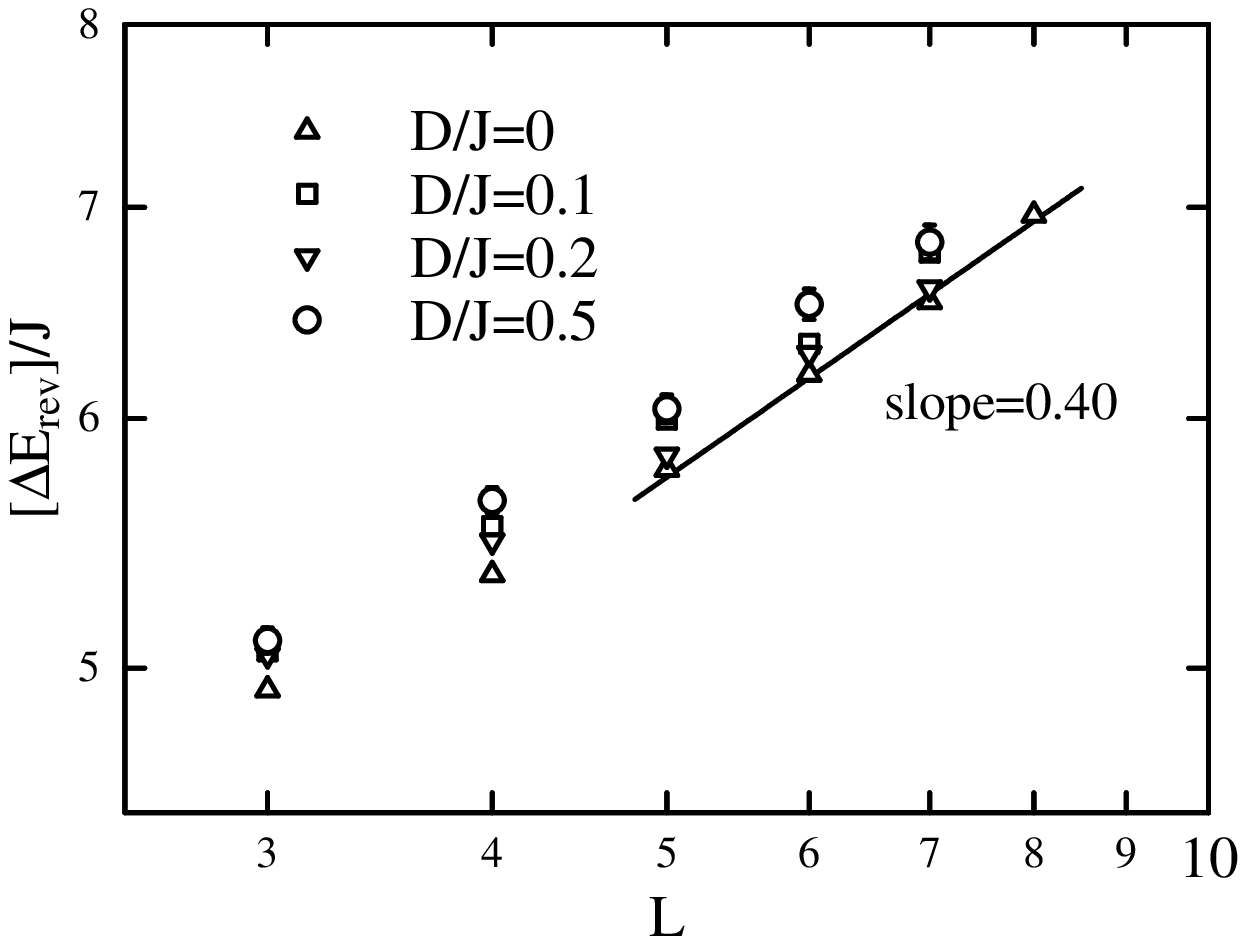}
\end{center}
  \caption{The lattice size dependence of the excess energy 
           $[\Delta{E}_{\rm rev}]$ of the Heisenberg SG model 
           in $d = 3$.}
  \label{fig:3}
\end{figure}
%%%%%%%%%%%%%%%%%%%%%%%%%%%%%%%%%%%%%%%%%%%%%%%%

Our results suggest $\theta_{\rm rev}, \theta_{\rm rot} > 0$ for both 
$D = 0$ and $D \ne 0$. However, their values are somewhat different 
with each other. 
It should be pointed out, however, that these values of $\theta_{\rm rev}$ 
and $\theta_{\rm rot}$ may vary for $L \rightarrow \infty$, 
because in the lattice size range studied here $[\Delta E_{\rm rot}(\pi/2)]$ 
is smaller than $[\Delta E_{\rm rev}]$ 
and the former increases more rapidly than the 
latter ($\theta_{\rm rot} > \theta_{\rm rev}$), then as $L$ increases 
further they would come close with each other. The convincing values of 
$\theta_{\rm rot}$ and $\theta_{\rm rev}$ would be given in that range of $L$. 
Unfortunately, the lattice sizes are still small to examine this speculation. 
Any way, both the analyses of $[\Delta E_{\rm rot}(\pi/2)]$ and 
$[\Delta E_{\rm rev}]$ suggest that the system tends to be {\it rigid} 
as the size of the lattice becomes larger. 
Note that we have also calculated the defect energies of the system for 
$D = 0$ using two replica boundary conditions\cite{Ozeki} 
and found that they also increase with similar, positive slopes of 
$\theta^{(\rm rep)}_{\rm rot} \sim 0.59$ and 
$\theta^{(\rm rep)}_{\rm rev} \sim 0.46$ for the $\pi$ rotation around 
the $z$-axis and the reversion, respectively.\cite{Endoh2} 
Hence, we conclude that {\it the defect energy method never 
gives evidence of the phase transition at $T_{\rm SG} = 0$}.

%%%%%%%%%%%%%%%%%%%%%%%%%%%%%%%%%%%%%%%%%%%%%%%%%%%%%%%%%%%%%%%%%%%%
\begin{figure}
\epsfxsize=0.8\linewidth
\begin{center}
  \epsfbox{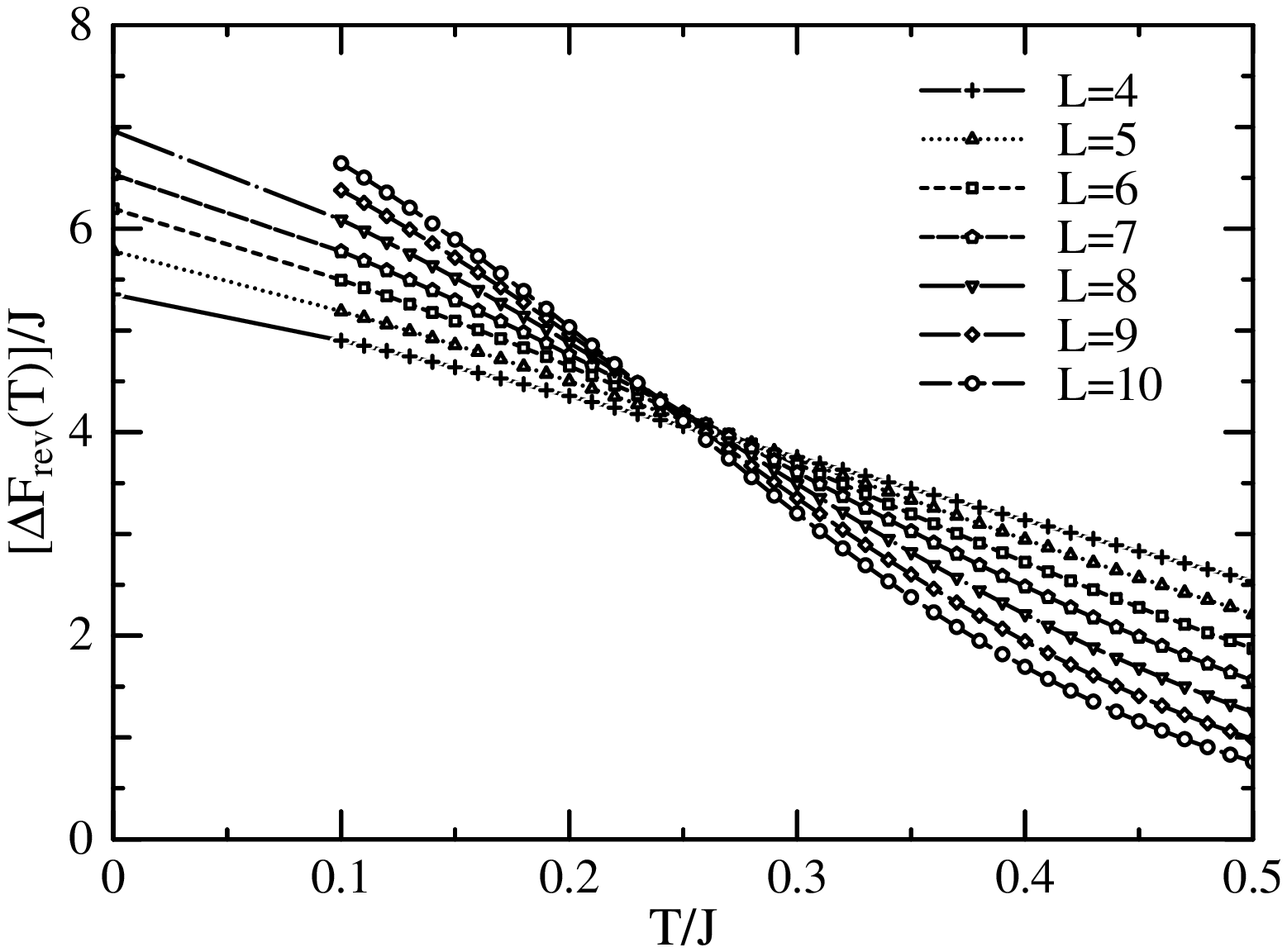}
\end{center}
  \caption{The lattice size dependence of the excess free energy 
         $[\Delta {F}_{\rm rev}(T)]$ of the Heisenberg SG model 
         for $D = 0$ in $d = 3$.\cite{Endoh2} Data at $T = 0$ 
         are $[\Delta {E}_{\rm rev}]$ presented in Fig.3
          }
  \label{fig:4}
\end{figure}
%%%%%%%%%%%%%%%%%%%%%%%%%%%%%%%%%%%%%%%%%%%%%%%%%%%%%%%%
What is the SG phase transition temperature $T_{\rm SG}$? 
We may estimate $T_{\rm SG}$ by calculating the excess free energy 
$[\Delta F_{\rm rot}(T)]$ and $[\Delta F_{\rm rev}(T)]$\cite{Ueno} 
at finite temperatures. 
In fact, we have also calculated these quantities for $D = 0$.\cite{Endoh2} 
The result of $[\Delta F_{\rm rev}(T)]$ is shown in Fig. 4. 
It is seen that, at high temperatures $[\Delta{F}_{\rm rev}(T)]$ decreases 
with increasing $L$, whereas at low temperatures they increase with $L$. 
One estimates the phase transition temperature from the crossing 
temperature of the free energies for various lattice sizes $L$.
In the present model, the crossing temperature $T_L$ for the lattice 
sizes $L$ and $L+1$ shifts systematically to the low temperature side with 
increasing $L$.
Then, we assumed that $T_L$ decreases linearly with $1/L$, and estimated 
$T_L$ for $L \rightarrow \infty$ as $T_{\infty}/J = 0.188 \pm 0.015$. 
Note that the same extrapolation for $[\Delta{F}_{\rm rot}(T)]$ gave 
$T_{\infty}/J = 0.192 \pm 0.015$. 
Therefore we may conclude that, if the SG phase transition occurs, 
the transition temperature is $T_{\rm SG} \sim 0.19J$.

%%%%%%%%%%%%%%%%%%%%%%  Section 3 %%%%%%%%%%%%%%%%%%%%%%%%%%%%%%%%%%%%%
\section{Monte Carlo Simulation}

Now we reexamine the SG phase transition itself. 
Here we consider the model on a simple cubic lattice of $L \times L 
\times (L+1) ( \equiv N) $ with skew boundary conditions along two $L$ 
directions and a periodic boundary condition along the $(L+1)$ direction.
We perform a MC simulation of the two-replica systems of the spins 
$\{\bm{S}_i\}$ and $\{\bm{T}_i\}$ using an exchange MC 
algorithm\cite{ExchngeM}.
We calculate the order-parameter probability distribution $P_L(q)$ of 
\begin{eqnarray}
	P_L(q) = [\langle\delta (q - Q)\rangle],
\end{eqnarray}
where $\langle \cdots \rangle$ and [$\cdots$] mean the thermal average 
and the bond distribution average, respectively. 
Here $Q$ is the spin overlap defined by 
\begin{eqnarray}
	Q = \sqrt{\frac{1}{3}\sum_{\alpha,\beta} (q^{\alpha\beta})^2},
\end{eqnarray}
with $q^{\alpha\beta} \equiv \frac{1}{N} \sum_{i=1}^N S_i^\alpha 
T_i^\beta$.
Using $P_L(q)$, we obtain two conventional SG quantities, i.e., 
the SG susceptibility $\chi_{\rm SG}$ 
and the Binder parameter $g(L,T)$ which are defined by 
\begin{eqnarray} 
   \chi_{\rm SG} &=& 3N[\langle q^2\rangle],\\  
	g(L,T) &=& \frac{1}{2} (11 - 9 \frac{[\langle q^4\rangle]}
			{[\langle q^2\rangle]^2}),
\end{eqnarray}
where $[\langle q^n \rangle] = \int q^n P_L(q) dq$. 
We examine the size and temperature dependences of these quantities 
both for $D = 0$ and for $D \neq 0$. 
The linear sizes of the lattice studied here are $L = 5 \sim 19$. 
Equilibration is checked by monitoring the stability of the results 
against at least two-times longer runs. 
The numbers of the samples are 480 for $L=5 \sim 9$, 192 for $L=11$,
96 for $L=15$, and 48 for $L=19$.

%%%%%%%%%%%%%%%%%%%%%%%%%%%%%%%%%%%%%%%%%%%%%%%%%%%%%%%%%%%%%%%%%%%
%
\begin{figure}
\begin{center}
\epsfxsize=0.7\linewidth
\epsfbox{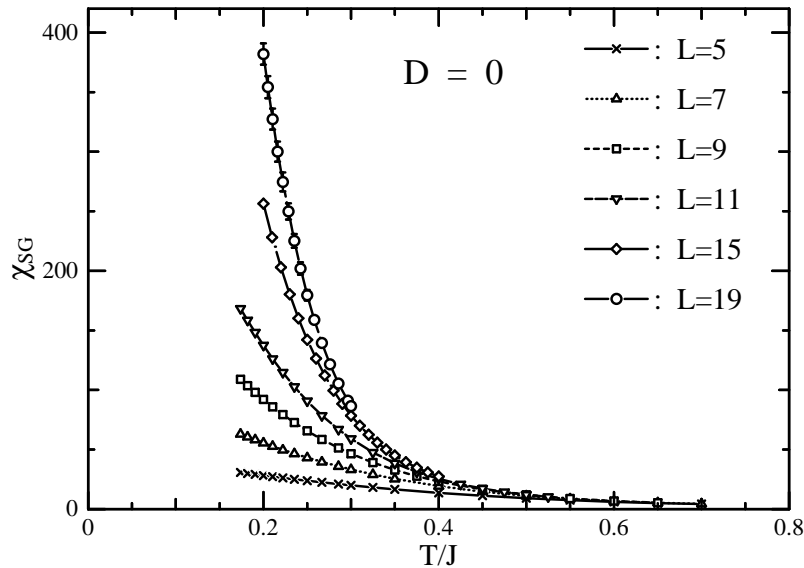}
%\epsfbox{MCFIG1.eps}
\end{center}
\caption{
Temperature dependences of the spin-glass susceptibility $\chi_{\rm SG}$
of the $\pm J$ Heisenberg model in $d=3$ for different sizes of the lattice
at $D = 0$. 
}
\label{MCfig:5}
\end{figure}

In Fig. 5, we show results of SG susceptibility $\chi_{\rm SG}$ of the 
model without the anisotropy ($D = 0$). As the temperature is decreased,
$\chi_{\rm SG}$ for larger $L$ increases rapidly. 
If the lower critical dimension $d_l$ is less than
the lattice dimension, $d_l < 3$, and the phase transition really
occurs at $T = T_{\rm SG}$, the data for different $L$ will be scaled as
\begin{eqnarray} 
   \chi_{SG} = L^{2-\eta}F(L^{1/\nu}(T-T_{\rm SG})),
\end{eqnarray}
where $\nu$ is the exponent of the correlation length and $\eta$ is 
the exponent which describes the decay of the correlation function at 
$T = T_{\rm SG}$. 
The scaling plots obtained by assuming $T_{\rm SG} \neq 0$ and 
$T_{\rm SG} = 0$ are shown in Fig. 6. 
The scaling with $T_{\rm SG} \neq 0$ works better than that with 
$T_{\rm SG} = 0$, even if the data for the smallest size $L = 5$ 
are ignored in the latter.\cite{Comm_gL}
Note that in the previous scaling analysis\cite{Iyota2}, 
$T_{\rm SG} = 0$ was estimated using the data for lattice of $L = 7 - 15$. 
Here, we use the data for a wider temperature range and add the data of 
the bigger lattice of $L = 19$. 
The phase transition temperature and the values of the critical exponent 
estimated here are $T_{\rm SG}/J = 0.18 \pm 0.01$, 
$\nu = 0.97 \pm 0.05$ and $\eta = -0.1 \pm 0.1$. 
We should emphasize that this value of $T_{\rm SG}$ is in good agreement 
with that estimated from the excess free energy of $T_{\rm SG}/J \sim 0.19$. 
It is noted, however, that the possibility of $T_{\rm SG} = 0$ is not 
ruled out from the scaling plot of Fig. 6(b), because in that 
case the temperature range of $T \gtrsim 0.2J$ would be out of a critical 
region.\cite{Moore} 
As the anisotropy is added, the transition temperature increases with $D$, 
i.e., $T_{\rm SG}/J = 0.32 \pm 0.03$ for $D = 0.2J$, and 
$T_{\rm SG}/J = 0.65 \pm 0.05$ for $D = 1.0J$.

%%%%%%%%%%%%%%%%%%%%%%%%%%%%%%%%%%%%%%%%%%%%%%%
\begin{figure}
\begin{center}
\epsfxsize=0.7\linewidth
\epsfbox{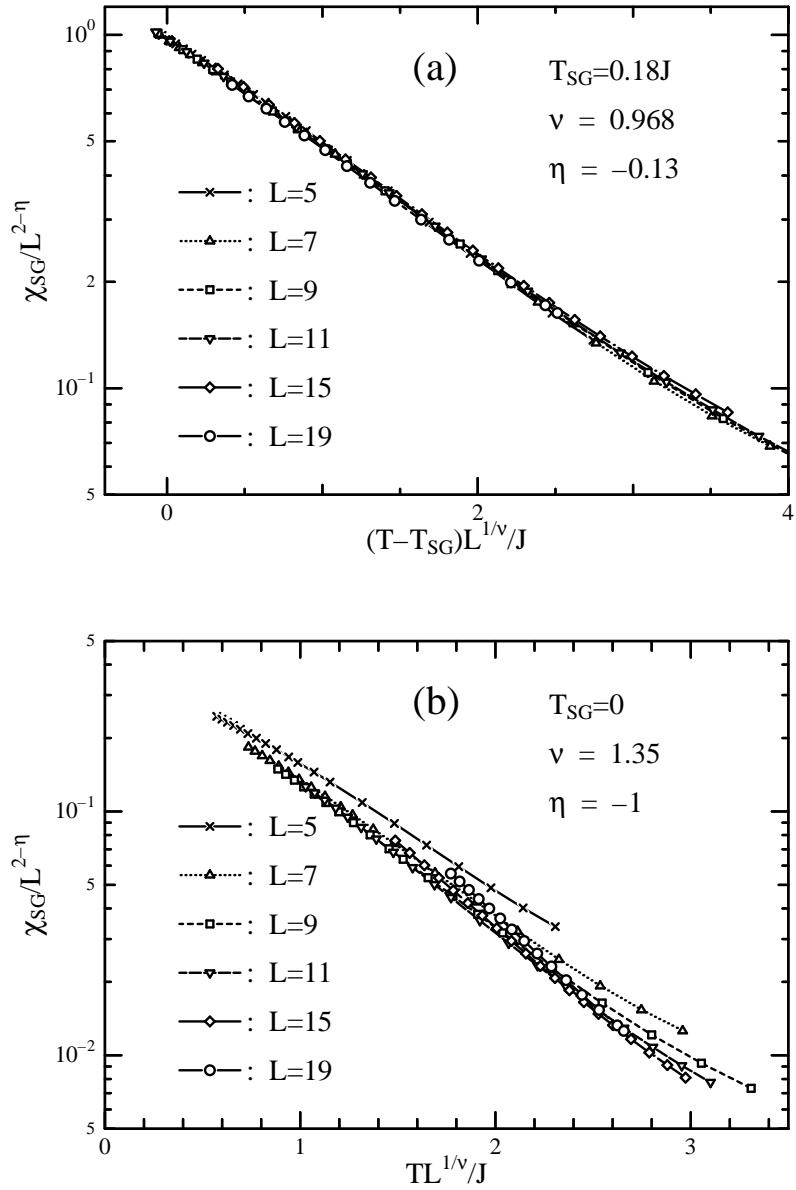}
\end{center}
\caption{ Typical examples of the finite size scaling plot of the SG 
susceptibility at $D = 0$ 
for (a) $T_{\rm SG} \neq 0$ and (b) $T_{\rm SG} = 0$. }
\label{fig:6}
\end{figure}
%%%%%%%%%%%%%%%%%%%%%%%%%%%%%%%%%%%%%%%%%%%%%

The Binder parameter $g(L,T)$ is the other quantity for examining the SG 
phase transition. 
It is believed that, if the SG phase transition occurs at $T_{\rm SG}$, 
$g(L,T)$'s for different $L$ cross at $T_{\rm SG}$. 
Contrary to our expectation, as shown in Fig. 7, they neither cross nor come 
together at $T_{\rm SG}$. This result seems to give the opposite view about 
the SG phase transition. 
However, the absence of the crossing of $g(L,T)$ was also seen in the $\pm J$ 
Heisenberg model in four dimensions ($d = 4$)\cite{Shira2} in which 
the SG phase transition is believed to occur at some finite temperature 
even when $D = 0$.\cite{Coluzzi} 
If the absence of the crossing of $g(L,T)$'s for finite $L$ says nothing 
about the SG phase transition, the same would be true when the anisotropy 
is present ($D \ne 0$). 
Then, we also calculate the Binder parameter of the model with $D \ne 0$. 
Here, since the system for $D \ne 0$ has the inversion symmetry, 
we also consider the spin overlap of the diagonal components for which, 
in eq. (4), 
\begin{eqnarray}
	Q_{\rm diag} = \frac{1}{N} \sum_{i=1}^N \bm{S}_{i}\bm{T}_{i} 
\end{eqnarray} 
is used instead of $Q$. 
Hereafter, we call the SG susceptibility and the Binder parameter 
calculated using $Q_{\rm diag}$ as the diagonal SG susceptibility and 
the diagonal Binder parameter and denote $\chi^{(\rm diag)}$ and 
$g^{(\rm diag)}(L,T)$, respectively.  
Results are presented in Figs. 8(a) and 8(b) for $g(L,T)$ and 
$g^{\rm (diag)}(L,T)$, respectively. 
In fact, $g(L,T)$'s for different $L$ neither cross nor come together. 
In contrast, $g(L,T)$ for larger $L$ exhibits a dip. 
As $D$ is increased, this property becomes more prominent. 
On the contrary, $g^{\rm (diag)}(L,T)$ exhibits {\it a usual behavior}. 
That is, as the temperature is decreased, $g^{\rm (diag)}(L,T)$ increases 
monotonically and its size dependence reverses. 
We suggest, hence, that the definition of the Binder parameter in terms of 
$Q_{\rm diag}$ is adequate for examining the phase transition 
for $D \ne 0$ and its crossing behavior supports the presence of 
the phase transition. 
It is noted, however, that the crossing temperature seems to deviate 
considerably from that estimated above. 
We think that this deviation comes from a finite size effect, because 
the crossing temperatures for different $L$'s exhibit a considerable 
$L$ dependence and, as $L$ increases, it seems to approach $T_{\rm SG}$.  
We have also calculated $g^{(\rm diag)}(L,T)$ in the 
case of $D = 0$ and found the absence of the crossing behavior. 
We think this difference in the behavior of $g^{(\rm diag)}(L,T)$ 
comes from the occurrence of the drift of the whole system 
due to the $O(3)$ symmetry for $D = 0$. 
In fact, the diagonal SG susceptibility $\chi^{(\rm diag)}_{\rm SG}$ 
for $D = 0$ have been found to be much smaller than $\chi_{\rm SG}$,  
whereas that for $D \ne 0$ is larger than $\chi_{\rm SG}$.\cite{Comm_Chi} 
We believe, hence, that the absence of the crossing of the usual 
Binder parameter $g(L,T)$ will not say the absence of the phase 
transition of this system. 
We speculate that, even when $D = 0$, if the system becomes free from the 
drift, $g^{(\rm diag)}(L,T)$ might exhibit a similar crossing behavior.

%%%%%%%%%%%%%%%%%%%%%%%%%%%%%%%%%
%
\begin{figure}
\epsfxsize=0.7\linewidth
\begin{center}
\epsfbox{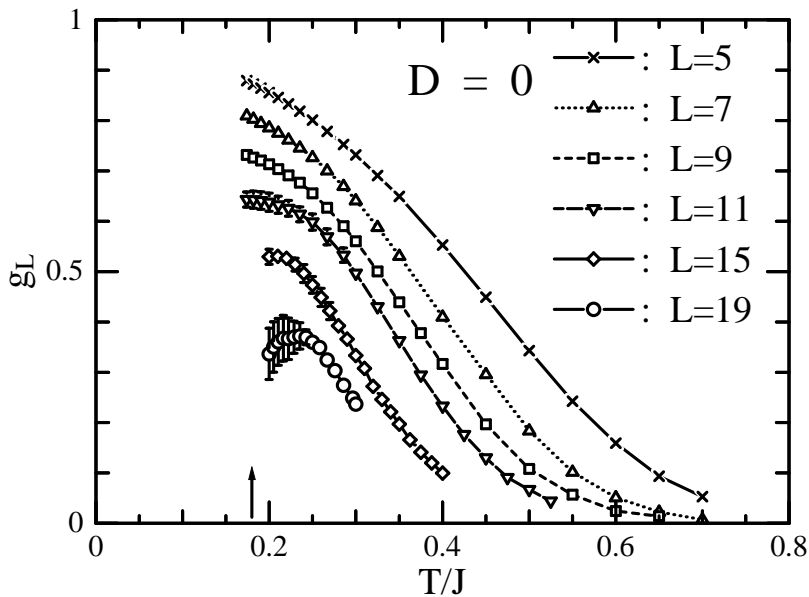}
\end{center}
\caption{
Temperature dependences of the Binder parameter $g(L,T)$ of the $\pm J$ 
Heisenberg model for $D = 0$ in $d=3$ for different sizes of the lattice. 
The arrow indicates the transition temperature $T_{\rm SG}$ estimated from 
the scaling plot of $\chi_{\rm SG}$ 
}
\label{fig:7}
\end{figure}
%%%%%%%%%%%%%%%%%%%%%%%%%%%%%%%%

%%%%%%%%%%%%%%%%%%%%%%%%%%%%%%%
\begin{figure}
\begin{center}
\epsfxsize=0.7\linewidth
\epsfbox{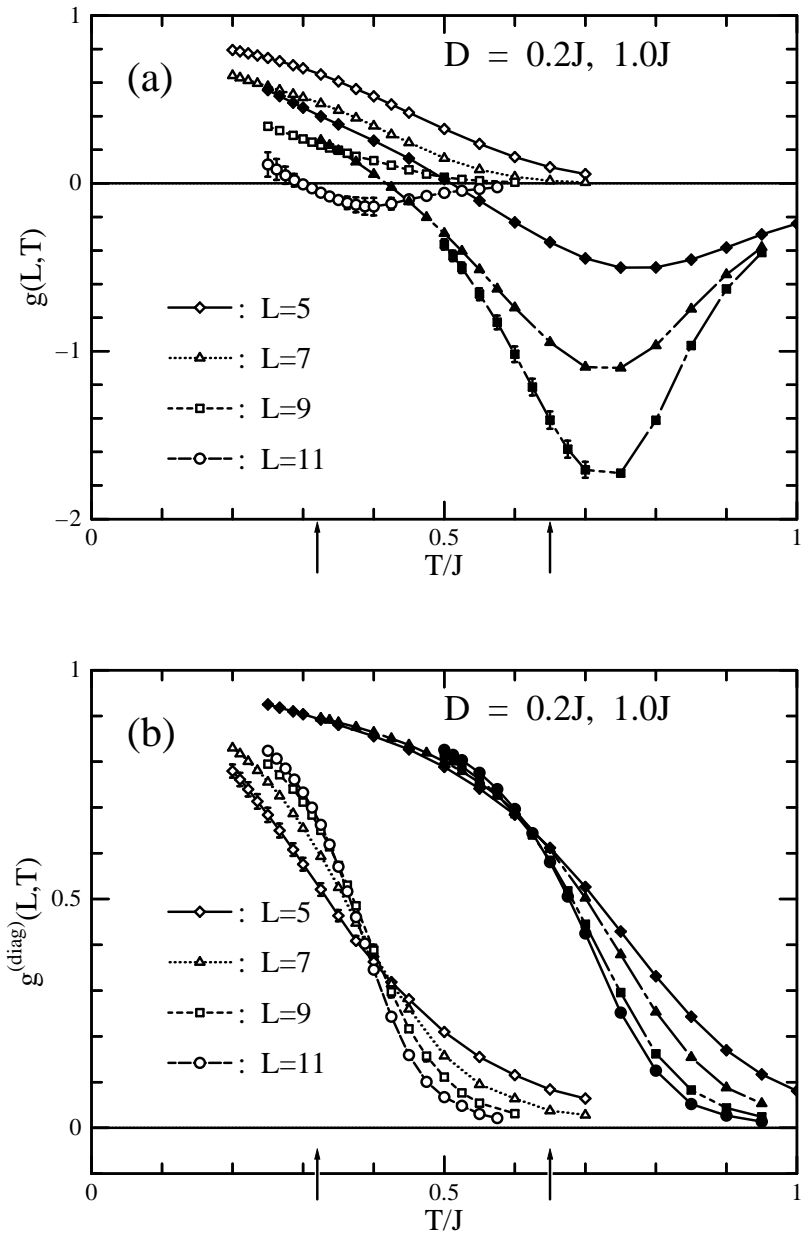}
\end{center}
\caption{
Temperature dependences of (a) the Binder parameter $g(L,T)$ 
and (b) the diagonal Binder parameter $g^{(\rm diag)}(L,T)$ of 
the $\pm J$ Heisenberg model in $d=3$ for different magnitude 
of the anisotropy $D$ and for different sizes of the lattice. 
Open symbols are for $D=0.2J$ and closed ones for $D=1.0J$. 
Arrows indicate the transition temperatures $T_{\rm SG}$ estimated from 
the scaling plot of $\chi_{\rm SG}$ 
} 
\label{fig:8}
\end{figure}
%%%%%%%%%%%%%%%%%%%%%%%%%%%%%%%%%

Recently, it was proposed that the quantities $A(L,T)$ and $G(L,T)$ that 
measure the order-parameter fluctuations (OPF) exhibit the crossing 
behavior at $T_{\rm SG}$ even if $g(L,T)$ does 
not:\cite{Hukushima1,Picco,Marinari2}
\begin{eqnarray}
     A(L,T) &=& \frac{[\langle q^2\rangle^2] - [\langle q^2\rangle]^2}
            {[\langle q^2\rangle]^2},   \\
    G(L,T) &=& \frac{[\langle q^2\rangle^2] 
     - [\langle q^2\rangle]^2}{[\langle q^4\rangle] - [\langle q^2\rangle]^2}.
\end{eqnarray}
Then we also calculate $A(L,T)$ and $G(L,T)$ and examine their 
$L$-dependences. 
In Fig. 9, we show $G(L,T)$ for different $L$ at $D = 0$ and $D = 0.2J$. 
When $D = 0$, $G(L,T)$'s for large $L(\geq 9)$ seem to come together near
$T_{\rm SG}$. This property becomes more prominent in the anisotropic case 
of $D = 0.2J$ where the data for smaller $L(= 5, 7)$ join. 
We have also seen that $A(L,T)$ for both $D = 0$ and $D \ne 0$ show 
a somewhat different crossing behaviors at a temperature a little higher 
than $T_{\rm SG}$.

%%%%%%%%%%%%%%%%%%%%%%%%%%%%%%%%%%%%%%%%%%%%%
\begin{figure}
\begin{center}
\epsfxsize=0.7\linewidth
\epsfbox{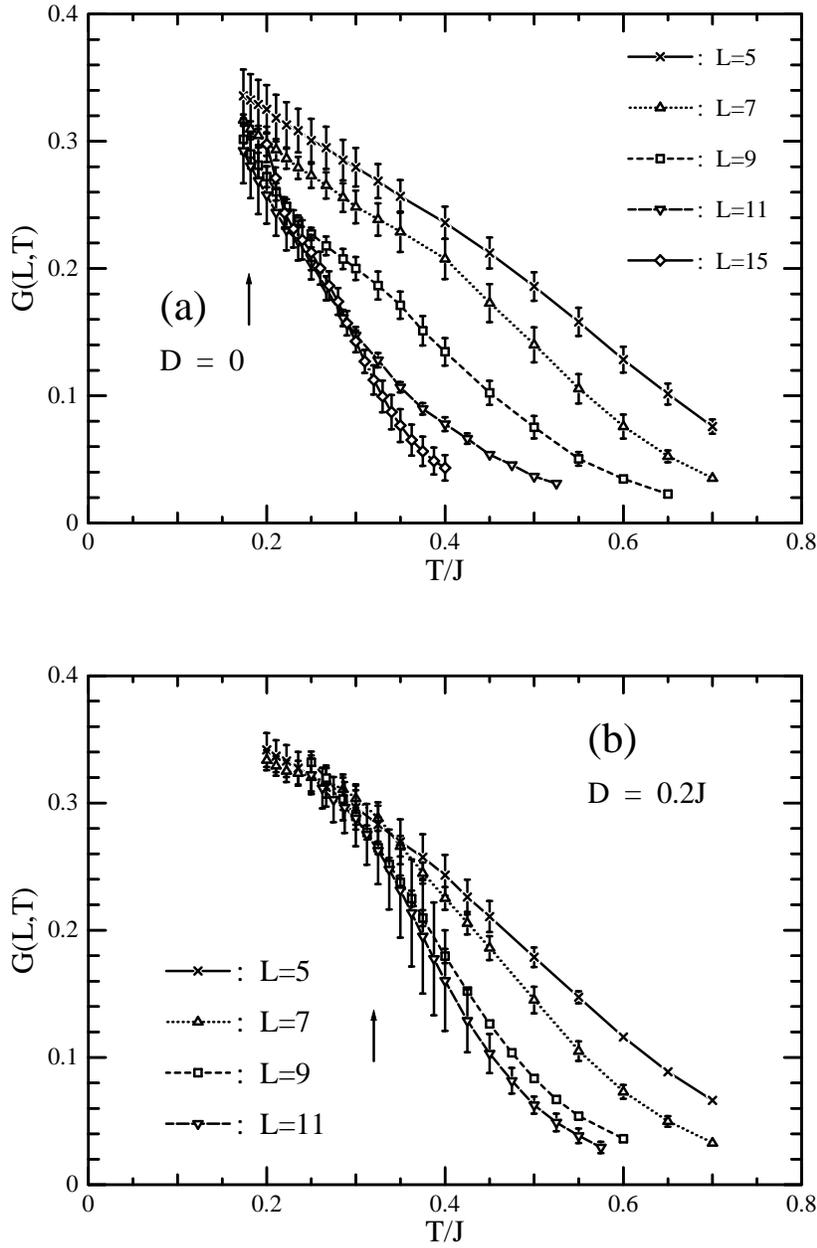}
\end{center}
\caption{
Temperature dependences of the order-parameter fluctuations $G(L,T)$ 
of the $\pm J$ Heisenberg model in $d=3$ for different sizes of the lattice 
: (a) $D = 0$, and (b) $D = 0.2J$ . 
The arrow indicates the transition temperature $T_{\rm SG}$ estimated from 
the scaling plot of $\chi_{\rm SG}$  }
\label{fig:9}
\end{figure}

%%%%%%%%%%%%%%%%%%%%%%%%%%%%%%%%%%%%%%%%%%%%

\section{Conclusion}

We have reexamined the spin-glass (SG) phase transition of the $\pm J$ 
Heisenberg models with and without the random anisotropy $D$ in three 
dimensions ($d = 3$). Attentions have been paid on the results of 
(i) the defect energy method and (ii) the Monte-Carlo method, 
because the evidences of the absence of the SG phase transition at 
a finite temperature have been given by these two methods. 
Our results have been summarized as follows. 

(i) The stiffness exponent $\theta$: We have shown that the previous 
result of $\theta < 0$ is not convincing because of the two reasons, i.e., 
(a) the meaning of the defect energy $[|\Delta E_L|]$ in the conventional 
method is not clear, and (b) even if the method is meaningful, 
the plot of $[|\Delta E_L|]$ as a function of $L$ curves considerably. 
We have proposed a new method which considers the stiffness of the 
lattice itself. By using the method, we have shown that $\theta > 0$ for 
$D \ne 0$ and the same is true for $D = 0$. Having considered the stiffness 
at finite temperatures, we have obtained the SG phase transition temperature 
$T_{\rm SG} \sim 0.19J$ for $D = 0$.  

(ii) The Monte Carlo method:  A large scale simulation have enabled us to 
make a scaling plot of the SG susceptibility $\chi_{\rm SG}$ which suggests 
the finite transiton temperature of $T_{\rm SG} \sim 0.18J$ for $D = 0$. 
The quantities $G(L,T)$ and $A(L,T)$ that measure the order-parameter 
fluctuations have exhibited a merging behavior near $T_{\rm SG}$, 
but the Binder parameter $g(L,T)$ has not exhibited the usual crossing 
behavior. However, analyses of the model with $D \ne 0$ have  
suggested that the absence of the crossing of $g(L,T)$'s 
will not mean the absence of the SG order in this model.

Our results have focused that, in contrary to the common belief, 
the SG phase transition occurs at a finite temperature. 
It should be noted that the two different method have given almost the 
the same SG transition temperature of $T_{\rm SG} \sim 0.19J$. 
This value of the transition temperature is very close to that 
estimated from nonequilibrium properties of the model, i.e., 
$T_{\rm SG} \sim 0.19J$ from the aging effect of the spin autocorrelation 
function\cite{Rot} and  $T_{\rm SG} \sim 0.21J$ from the nonequilibrium 
relaxation method\cite{Nakamura}. 
Quite recently, Lee and Young\cite{Lee} studied the Gaussian 
Heisenberg model, and suggested $T_{\rm SG}/J = 0.16 \pm 0.02$. 
This value of the transition temperature is also reasonably close to our 
value, considering the difference of the bond distribution. 
Hence we conclude that the model exhibits the SG phase transition 
even when the anisotropy is absent and its transition 
temperature is $T_{\rm SG} \sim 0.19J$.

We give two comments. One might think that, for a larger $D$, the Ising 
values of the exponents ($\theta \sim 0.2$\cite{BrayM}, and $\nu \sim 2.$ 
and $\eta = -0.3$\cite{Ballesteros}) should be recovered. 
However, we consider that this opinion is not necessarily true, 
because the random anisotropy in the model of eq.(1) is not uniaxial. 
Of course, we could not rule out the possibility that a finite-size effect 
masks true values. 
The other comment is that the strange behavior of $g(L,T)$ of the 
Heisenberg SG model will come from the choice of the order parameter. 
When $D \neq 0$, the order parameter should be chosen as the sum of 
the diagonal components of the spin overlap, because only the inversion 
symmetry exists. 
We speculate that the same will be true in the isotropic case of $D = 0$, 
though the $O(3)$ symmetry recovers. To examine this speculation, we 
are currently making the simulation removing the uniform rotation of the 
system.\cite{Rot}

%%%%%%%%%%%%%%%%%%%%%%%%%%%%%%%%%%%%%%%%%%%%%%%%%%%%%%%%%%%%%
%%%%%%%%%%%%%%%%%%%%%%%%%%%%%%%%%%%%%%%%%%%%%%%%%%%%%%%%%%%%%

\bigskip

The authors would like to thank Dr. T. Nakamura for his valuable 
discussions.

%%%%%%%%%%%%%%%%%%%%%%%%%%%%%%%%%%%%%%%%%%%%%%%%%%%%%%%%%%%%%%%%%%%


\begin{thebibliography}{99}
\bibitem{Comm_gL} In the case of $T_{\rm SG} = 0$, we have made the scaling 
         plot assuming $\eta$ being an adjustable parameter and found 
         that quality of the scaling plot is almost the same in the range of 
         $-1.0 \leq \eta \leq -0.9$.
\bibitem{Comm_Chi} We have also made the scaling plot of 
         $\chi^{(\rm diag)}_{\rm SG}$ and found that data are scaled fairly 
         well using a similar transition temperature for both $D = 0$ and 
         $D \neq 0$.
\end{thebibliography}
\end{document}